\documentclass[twocolumn,A4]{article}
\usepackage[dvips]{graphics}
\begin{document}
\onecolumn
\begin{center}
{\bf{\Large Quantum Transport in a Biphenyl Molecule: Effects of
Magnetic Flux}}\\
~\\
Santanu K. Maiti$^{\dag,\ddag,*}$ \\
~\\
{\em $^{\dag}$Theoretical Condensed Matter Physics Division,\\
Saha Institute of Nuclear Physics, \\
1/AF, Bidhannagar, Kolkata-700 064, India \\
$^{\ddag}$Department of Physics, Narasinha Dutt College,
129, Belilious Road, Howrah-711 101, India} \\
~\\
{\bf Abstract}
\end{center}
Electron transport properties of a biphenyl molecule are studied based on
the Green's function formalism. The molecule is sandwiched between two 
metallic electrodes, where each benzene ring is threaded by a magnetic 
flux $\phi$. The results are focused on the effects of the molecule to 
electrode coupling strength and the magnetic flux $\phi$. Our numerical 
study shows that, for a fixed molecular coupling, the current amplitude 
across the bridge can be {\em regulated} significantly just by tuning 
the flux $\phi$. This aspect may be utilized in designing nano-scale 
electronic circuits.  
\vskip 1cm
\begin{flushleft}
\end{flushleft}
\vskip 4.5in
\noindent
{\bf ~$^*$Corresponding Author}: Santanu K. Maiti

Electronic mail: santanu.maiti@saha.ac.in
\newpage
\twocolumn

\section{Introduction}

In the last few decades, electron transport properties through nano-scale 
systems$^{1-4}$ have been studied enormously, and recently people are really 
getting very excited in fabrication of electronic circuits by using single 
molecule or cluster of molecules$^{5-6}$ due to their unique properties. 
In the very early days, Aviram and Ratner$^7$ first predicted theoretically 
the electron transport through a molecular system, and later many 
experiments$^{8-12}$ were performed to understand the basic mechanism 
underlying such transport. In a molecular bridge, the electron transmission 
is controlled by several important factors and all these effects have to be 
taken into account properly. The most important issue is probably the 
quantum interference effects$^{13-16}$ among the electron waves traversing
through different arms of the molecule. Another important issue is the
molecular coupling to the side attached electrodes.$^{17}$ Tuning 
this coupling, one can control the current amplitude very nicely across 
the bridge. Similar to these, there are several other factors like the 
geometry of the molecule,$^{18}$ electron-electron correlations,$^{19}$ 
dynamical fluctuations,$^{20-21}$ etc., which provide rich effects in 
the electron transport.

In the present article we focus on the electron transport properties of
a biphenyl molecule, where each benzene ring is threaded by a magnetic 
flux $\phi$, the so-called Aharonov-Bohm (AB) flux. Quite interestingly 
we see that, keeping all the other parameters as invariant, the current 
amplitude across the molecule can be regulated very nicely simply by 
tuning the magnetic flux $\phi$. Thus we can design an electronic circuit 
by using the biphenyl molecule and the electron transmission through the 
circuit can be regulated efficiently just by controlling the parameter 
$\phi$. This phenomenon can be utilized in designing the future 
nano-electronic circuits. Here we provide a very simple analytical 
formulation of the transport problem through the biphenyl molecule using 
the tight-binding Hamiltonian, and the coupling of the molecule to the 
side attached electrodes is treated through the Newns-Anderson 
chemisorption theory.$^{22-24}$ There exist several {\em ab initio} 
methods$^{25-29}$ as well as model calculations$^{23-24,30-31}$ to 
determine the molecular conductance. For our illustrative purposes, 
here we concentrate on the model calculations, since the attention 
is drawn only on the qualitative behavior of the physical quantities 
rather than the quantitative study.

The paper is organized as follow. Following the introduction (Section $1$),
in Section $2$, we present the model and the theoretical formulations for 
our calculations. Section $3$ discusses the significant results, and finally, 
we summarize our results in Section $4$.

\section{Model and the synopsis of the theoretical background}

\begin{figure}[ht]
{\centering \resizebox*{7cm}{1.75cm}{\includegraphics{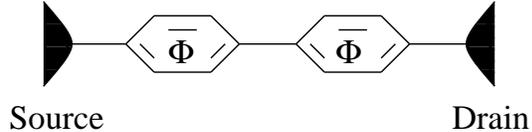}}\par}
\caption{(Color online). Schematic view of a biphenyl molecule attached 
to two electrodes, where each benzene molecule is threaded by a magnetic 
flux $\phi$.}
\label{biphenyl}
\end{figure}

Let us refer to Fig.~\ref{biphenyl}. A biphenyl molecule is attached to
two metallic electrodes, namely, source and drain, where each benzene 
ring is threaded by a magnetic flux $\phi$. In the actual experimental 
setup, gold (Au) electrodes are used, and, the molecule is coupled to 
the electrodes
through thiol (SH) groups in the chemisorption technique. Here we use
a simple tight-binding Hamiltonian to describe the biphenyl molecule,
and within the non-interacting picture this can be expressed in this
form,
\begin{equation}
H_M=\sum_i \epsilon_i c_i^{\dagger} c_i + \sum_{<ij>} t \left(
e^{i \theta} c_i^{\dagger} c_j + e^{-i \theta} c_j^{\dagger} c_i\right)
\label{equ4}
\end{equation}
where $\epsilon_i$ and $t$ represent the on-site energy and the 
nearest-neighbor ($j=i\pm 1$) hopping strength, respectively. 
$c_i^{\dagger}$ ($c_i$) corresponds to the creation (annihilation) 
operator of an electron at the site $i$. In this expression, 
$\theta=2 \pi \phi/N$ is the phase factor due to the flux $\phi$ (measured
in units of $\phi_0=ch/e$, the elementary flux quantum), where $N$ 
corresponds to the total number of carbon-type sites in each benzene ring. 
A similar type of tight-binding Hamiltonian is also used for the description 
of the electrodes, where the site energy and the nearest-neighbor hopping 
strength of the electrodes are represented by the parameters $\epsilon_0$ 
and $v$, respectively. The molecule is coupled to the electrodes by the 
parameters $\tau_S$ and $\tau_D$, where they (coupling parameters) 
correspond to the coupling strengths with the source and drain, respectively.

At much low temperatures and bias voltage, we use the Landauer conductance 
formula$^{32-33}$ to calculate the conductance $g$ of the molecule
which can be expressed as,
\begin{equation}
g=\frac{2e^2}{h} T
\label{equ1}
\end{equation}
where the transmission probability $T$ becomes,$^{32-33}$
\begin{equation}
T=Tr\left[\Gamma_S G_M^r \Gamma_D G_M^a\right]
\label{equ2}
\end{equation}
In this expression $G_M^r$  and $G_M^a$ correspond to the retarded and 
advanced Green's functions of the molecule, and $\Gamma_S$ and $\Gamma_D$ 
describe the couplings of the molecule with the source and drain, 
respectively. The Green's function of the molecule is in the form,
\begin{equation}
G_M=\left(E-H_M-\Sigma_S-\Sigma_D\right)^{-1}
\label{equ3}
\end{equation}
where $E$ is the energy of the injecting electron and $H_M$ represents
the molecular Hamiltonian as prescribed above in Eq.~(\ref{equ4}). The
parameters $\Sigma_S$ and $\Sigma_D$ represent the self-energies due to 
the molecular coupling with the source and drain, respectively, where all
the information of this coupling are included into these self-energies
and are described by the Newns-Anderson chemisorption theory.$^{22-24}$ 
Using the Newns-Anderson model, we can directly study the conductance 
in terms of the properties of the molecular electronic structure.

The current passing through the molecule can be regarded as a single 
electron scattering process between the two reservoirs of charge 
carriers. The current-voltage relationship can be obtained from the 
expression,$^{32}$
\begin{equation}
I(V)=\frac{e}{\pi \hbar}\int_{-\infty}^{\infty} \left(f_S-f_D\right) 
T(E) dE
\label{equ5}
\end{equation}
where $f_{S(D)}=f\left(E-\mu_{S(D)}\right)$ gives the Fermi distribution
function with the electrochemical potential $\mu_{S(D)}=E_F\pm eV/2$.
Usually, the electric field inside the molecule, especially for small 
molecules, seems to have a minimal effect on the $g$-$E$ characteristics. 
Thus it introduces very little error if we assume that, the entire voltage is
dropped across the molecule-electrode interfaces. The $g$-$E$ characteristics
are not significantly altered. On the other hand, for larger molecules
and higher bias voltage, the electric field inside the molecule may play a
more significant role depending on the size and the structure of the
molecule,$^{34}$ though the effect becomes quite small.

All the results described in the present communication are performed at 
absolute zero temperature. However, they should remain valid even in a 
certain range of finite temperature ($\sim 300$ K). This is due to the 
fact that the broadening of the energy levels of the molecule due to the 
molecule-electrode coupling is, in general, much larger than that of 
the thermal broadening.$^{32}$ For simplicity, we take the unit 
$c=e=h=1$ in our present calculations.

\section{Results and discussion}

To illustrate the results, let us first mention the values of the different 
parameters used for the numerical calculations. In the biphenyl molecule, 
the nearest-neighbor hopping strength ($t$) alternates between the two 
values,$^{35}$ those are respectively taken as 
\begin{figure}[ht]
{\centering \resizebox*{8cm}{9cm}{\includegraphics{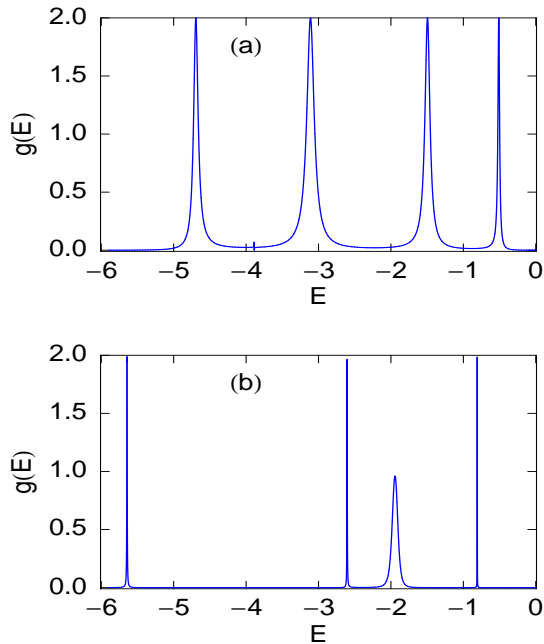}}\par}
\caption{(Color online). $g(E)$-$E$ curves for the biphenyl molecule in 
the weak-coupling limit, where (a) $\phi=0$ and (b) $\phi=0.45$.}
\label{weakcond}
\end{figure}
$-2.55$ and $-2.85$. The on-site energy ($\epsilon_i$) corresponding to 
carbon-type sites in this molecule is set to $-6.6$.$^{35}$ On the 
other hand, for the side attached electrodes the on-site energy 
($\epsilon_0$) and the nearest-neighbor hopping strength ($v$) are fixed 
to $0$ and $3$, respectively. The Fermi energy $E_F$ is set to $0$.
Throughout the study, we focus our results
for the two limiting cases depending on the strength of the molecular 
coupling with the source and drain. Case $I$: The weak-coupling limit. 
It is described by the condition $\tau_{S(D)} << t$. For this regime
we choose $\tau_S=\tau_D=0.75$. Case $II$: The strong-coupling limit.
This is specified by the condition $\tau_{S(D)} \sim t$. In this
particular regime, we set the values of the parameters as $\tau_S=\tau_D
=2.5$.

In Fig.~\ref{weakcond}, we display the conductance $g$ of the biphenyl
molecule as a function of the energy $E$ in the limit of weak-coupling.
Figure~\ref{weakcond}(a) corresponds to the conductance spectrum of the 
biphenyl molecule in the absence of any magnetic flux $\phi$. The
conductance spectrum shows very sharp resonant peaks for some particular
energy values, while for all other energies, $g$ becomes almost zero.
\begin{figure}[ht]
{\centering \resizebox*{8cm}{9cm}{\includegraphics{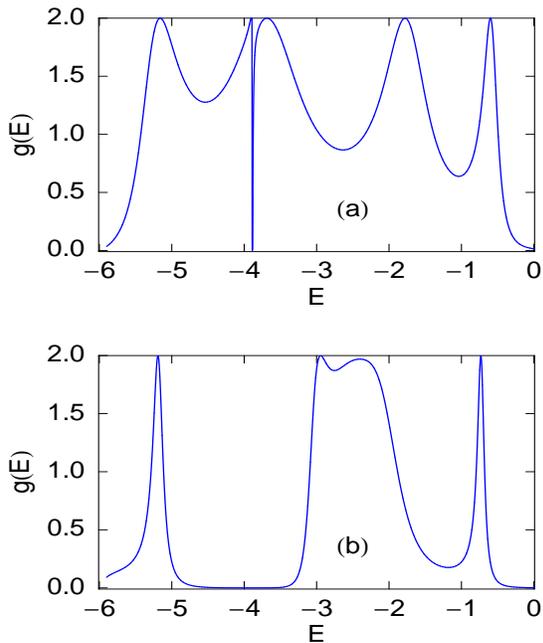}}\par}
\caption{(Color online). $g(E)$-$E$ curves for the biphenyl molecule in the 
strong-coupling limit, where (a) $\phi=0$ and (b) $\phi=0.3$.}
\label{strongcond}
\end{figure}
At these resonances, $g$ approaches the value $2$, and therefore, the
transmission probability $T$ goes to unity since we have the relation
$g=2T$ from the Landauer conductance formula (see Eq.~(\ref{equ1})).
These conductance peaks are associated with the molecular energy levels.
Thus the conductance spectrum manifests itself the energy spectrum of the
molecule. In the presence of the magnetic flux, the conductance spectrum 
gets modified significantly. Figure~\ref{weakcond}(b) shows the $g$-$E$ 
characteristics for the biphenyl molecule where we set $\phi$ to $0.45$. 
From this spectrum it is observed that the widths of the resonant peaks 
become much narrow compared to the case when $\phi=0$. Not only that, 
some resonant peaks also disappear and get reduced height in the presence 
of $\phi$. Here it is examined that, the width of the resonant peaks 
gradually decreases as the flux $\phi$ changes from $0$ to $\phi_0/2$, 
where $\phi_0=ch/e$ is the elementary flux-quantum. This can be implemented
as follow. In the bridge system the electron passes from the source to 
the drain. In this case the electron waves passing through the different
paths of the molecular rings may interfere constructively or destructively,
leading to stronger or weaker transmission probability across the molecule.
Applying the magnetic flux, the interference condition of the electron
waves among the different pathways can be controlled, and accordingly,
\begin{figure}[ht]
{\centering \resizebox*{8cm}{10cm}{\includegraphics{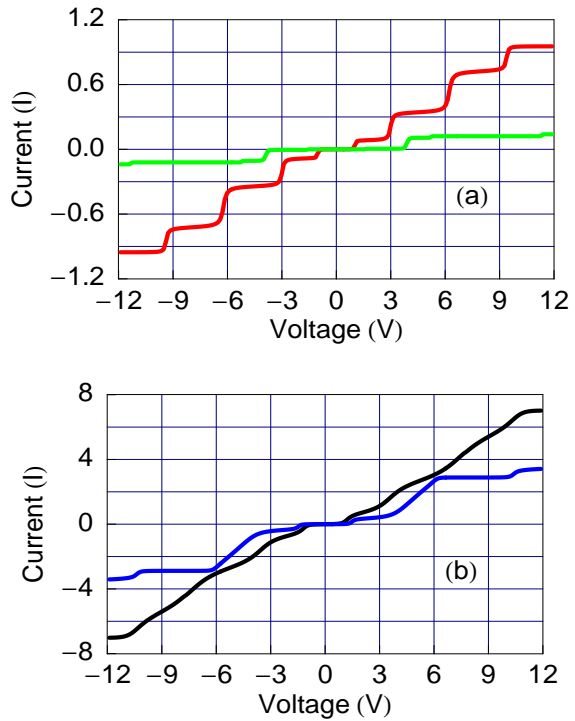}}\par}
\caption{(Color online). $I$-$V$ characteristics for the biphenyl molecule. 
(a) weak-coupling limit, where the red and green curves correspond to 
$\phi=0$ and $0.45$, respectively. (b) strong-coupling limit, where the 
black and blue curves correspond to $\phi=0$ and $0.3$, respectively.} 
\label{curr}
\end{figure}
the conductance spectrum gets modified. Thus quantum interference effect
plays an important role in the study of the electron transport in a
molecular bridge system. The effect of the molecular coupling is also very
interesting. To emphasize it, in Fig.~\ref{strongcond} we plot the $g$-$E$ 
characteristics of the biphenyl molecule in the limit of strong-coupling. 
Figure~\ref{strongcond}(a) represents the result for $\phi=0$, while 
Fig.~\ref{strongcond}(b) corresponds to the variation when $\phi$ is set 
to $0.3$. In this strong-coupling case, the width of the resonant peaks 
gets sufficient broadening compared to the weak-coupling limit. The 
contribution of such broadening comes from the 
imaginary parts of the self-energies $\Sigma_S$ and $\Sigma_D$.$^{32}$
It is important to note that the real parts of the self-energies provide
only the shift of the energy levels. Similar to the weak-coupling case,
here also the width of the resonant peaks becomes reduced with the flux
$\phi$. Another significant feature observed from the conductance spectra
is the existence of the anti-resonant state. Figure~\ref{strongcond}(a) 
shows that it appears across the energy $E=-3.9$, and for the weak-coupling 
limit it is also observed for the same energy (see Fig.~\ref{weakcond}(a)), 
but due to the smallness of the width it is not quite transparent like as 
in the case of strong molecular coupling. The anti-resonant states are 
specific to the interferometric nature of the molecule and do not appear 
in the conventional scattering problems like one-dimensional potential 
barriers, etc. 

The behavior of the electron transfer through the molecule becomes much more 
clearly visible by investigating the current-voltage characteristics, rather 
than the transmission spectra. The current $I$ is computed by the integration
procedure of the transmission function ($T$) (see Eq.~(\ref{equ5})), where 
the function $T$ varies exactly similar to the conductance spectra, differ
only in magnitude by a factor $2$, since the relation $g=2T$ holds from
the Landauer conductance formula (Eq.~(\ref{equ1})). The variation of the 
current-voltage characteristics of the biphenyl molecule in the limit of 
weak-coupling is shown in Fig.~\ref{curr}(a), where the red and green 
curves correspond to the results for $\phi=0$ and $0.45$, respectively. 
The current exhibits staircase like
structure with sharp steps as a function of the bias voltage $V$. This
is due to the existence of the sharp resonant peaks in the conductance
spectrum, since the current is computed by the integration procedure of the
transmission function $T$. With the increase of the applied bias voltage
$V$, the electrochemical potentials in the electrodes cross one of the
molecular energy levels and produce a jump in the $I$-$V$ characteristics.
The effect of $\phi$ is significantly observed from the green curve. The
current gets reduced to a large extent for this $\phi$ value. This can be
clearly understood from the conductance spectrum (see Fig.~\ref{weakcond}(b)) 
where the widths of the resonant peaks become much narrow compared to the 
case where $\phi=0$. The step-like behavior almost disappears in the strong 
molecular coupling limit. As illustrative example, in Fig.~\ref{curr}(b) we 
plot the $I$-$V$ characteristics for the molecule in the strong-coupling 
limit, where the black and blue curves represent the results for the cases 
$\phi=0$ and $0.3$, respectively. The current varies almost continuously with 
the bias voltage $V$ and achieves much higher amplitude than the 
weak-coupling limit. This phenomenon can be noticed clearly by observing 
the areas under the $g$-$E$ curves presented in Fig.~\ref{strongcond}. 
Here the current amplitude is also reduced with the flux $\phi$ (see the 
green curve).

\begin{figure}[ht]
{\centering \resizebox*{8cm}{5cm}{\includegraphics{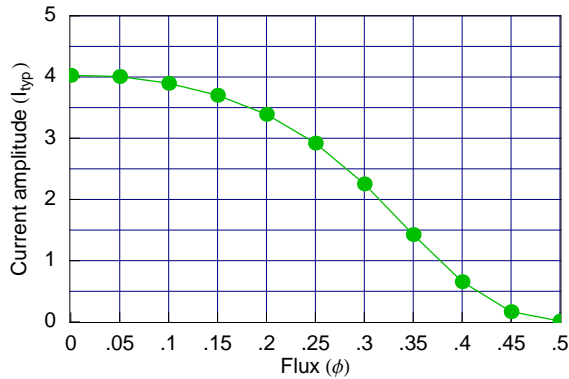}}\par}
\caption{(Color online). Variation of the typical current amplitude 
($I_{typ}$) as a function of the flux $\phi$ across the molecular bridge 
in the limit of strong molecular coupling.}
\label{strongcurramp}
\end{figure}

To emphasize the dependence of the current $I$ on the magnetic flux $\phi$, 
in Fig.~\ref{strongcurramp} we show the variation of the typical current 
amplitude ($I_{typ}$)
as a function of $\phi$. The typical current amplitude is computed from
the expression $I_{typ}=\sqrt{<I^2>_V}$, where the averaging over the bias 
voltage $V$ is done within the range $-12$ to $12$. Our result predicts
that the current amplitude smoothly decreases with the flux $\phi$ when
it changes from $0$ to the half flux-quantum value. This result is calculated
for the strong-coupling limit, and, the similar nature is also observed in
the case of weak-coupling. Thus we can say that, for a fixed molecular
coupling, the current amplitude can be controlled efficiently by tuning
the magnetic flux $\phi$.

\section{Concluding remarks}

In conclusion, we have used a parametric approach based on the tight-binding
model to investigate the electron transport properties of a biphenyl 
molecule attached to two metallic electrodes. In the molecule, each
benzene ring is threaded by a magnetic flux $\phi$ and the coupling of the
molecule to the electrodes have been described by the Newns-Anderson
chemisorption theory.$^{22-24}$ Here we have focused our results on the 
aspects of (a) the molecular coupling and (b) the magnetic flux $\phi$. 
Our numerical results have shown that, for a fixed molecular coupling 
strength, the current amplitude across the biphenyl molecule can be 
regulated significantly by controlling the flux $\phi$. This aspect may 
be utilized in designing a tailor made nano-scale electronic circuit.

This is our first step to describe how the electron transport in a biphenyl
molecule can be controlled very nicely by means of the magnetic flux $\phi$. 
Here we have used several realistic assumptions by ignoring the effects of 
the electron-electron correlation, disorder, etc. We need further study
in this particular system by incorporating all these effects.

\vskip 0.3in
\noindent
{\bf\Large Acknowledgments}
\vskip 0.2in
\noindent
I acknowledge with deep sense of gratitude the illuminating comments and
suggestions I have received from Prof. Arunava Chakrabarti and Prof. 
Shreekantha Sil during the calculations.

\end{document}